\documentclass[12pt]{article}
\usepackage{amsmath, amssymb, mathrsfs, bm, natbib}
\usepackage[usenames]{color}
\usepackage[dvips]{graphicx}
\usepackage{rotating}

\usepackage{setspace}
\usepackage{mathrsfs}
\usepackage{amssymb,amsmath,amsthm,mathrsfs,bm,verbatim}
\usepackage{indentfirst}
\usepackage{listings}
\usepackage{booktabs,multirow}

\newcommand{\mbs}[1]{\boldsymbol{#1}}

\begin{document}

\begin{center}
\Large{Efficient Global Monitoring Statistics for High-Dimensional Data}
\end{center}

\begin{center}
{\large Jun Li}\\
     Department of Statistics, University of California - Riverside
\end{center}

\begin{abstract}
\noindent
Global monitoring statistics play an important role for developing efficient monitoring schemes for high-dimensional data streams. A number of global monitoring statistics have been proposed in the literature. However, most of them only work for certain types of abnormal scenarios under specific model assumptions. How to develop global monitoring statistics that are powerful for any abnormal scenarios under flexible model assumptions is a long-standing problem in the statistical process monitoring field. To provide a potential solution to this problem, we propose a novel class of global monitoring statistics by making use of the quantile information in the underlying distribution of the local monitoring statistic. Our proposed global monitoring statistics are easy to calculate and can work under flexible model assumptions since they can be built on any local monitoring statistic that is suitable for monitoring a single data stream. Our simulation studies show that the proposed global monitoring statistics perform well across a broad range of settings, and compare favorably with existing methods.
\end{abstract}


\section{Introduction}\label{sec:intro}
Advanced manufacturing and data acquisition technologies have made the gathering of high-dimensional data possible in many fields. The demand for efficient online monitoring tools for such data has never been greater. Depending on the purpose, two types of monitoring schemes are needed.

The first type is for applications in which changes in any of data streams indicate the same abnormality in the whole system and require a single, uniform corrective action. As a result, those applications do not require the identification of abnormal data streams. For example, in the environmental monitoring application, hundreds or thousands of sensors are usually deployed to monitor certain environmental factors. The abnormality of data from any of those sensors will indicate a general abnormality in those environmental factors. In many cases, it is not of interest to identify which sensors have caused the alarm. Since it is not required to identify abnormal data streams for this type of monitoring scheme, a popular approach for developing monitoring schemes of this type is to use a single global monitoring statistic to track all data streams jointly. As a result, a global monitoring statistic that is powerful for detecting any abnormal scenario is the key for developing any efficient monitoring scheme of this type.

The second type of monitoring scheme is for applications in which changes in different data streams indicate different problems in the system, requiring unique corrective actions. As a result, this type of monitoring scheme needs to identify which data stream is experiencing abnormal activities. For example, in the network traffic surveillance application, the network traffic data from different data streams are associated with different IP addresses. When some abnormality in the data occurs, identifying which data stream or IP address has caused the problem is very important, as doing so helps pinpoint the cause and guides future corrective actions. As pointed out in Li (2017a), the following two-stage strategy can be very effective to develop this type of monitoring scheme. In the first stage, a global monitoring statistic is used to decide whether there is any abnormal data stream. If this is the case, the second stage is activated and a local monitoring statistic is used to decide which data streams are abnormal. As shown in Li (2017a), this two-stage strategy has better performance than the one-stage strategy. Based on this two-stage strategy, it is evident that what global monitoring statistic to use in the first stage becomes critical, since it will ultimately affect the effectiveness of the resulting monitoring scheme in terms of how quickly it will raise an alarm when some of the data streams start experiencing abnormal activities.

From the above discussions, it is clear that an efficient global monitoring statistic is important for developing efficient monitoring schemes of both types. How to construct efficient global monitoring statistics for high-dimensional data streams has been an active research topic in the statistical process monitoring field. A number of global monitoring statistics have been proposed. However, most of them only work for certain types of abnormal scenarios under specific model assumptions. For example, under the assumption that each data stream follows a normal distribution and using the cumulative sum (CUSUM) statistic as the local monitoring statistic for each data stream, Tartakovsky et al. (2006) and  Mei (2010) proposed using the maximum and sum of those CUSUM statistics, respectively, as the global monitoring statistic. It has been shown that  the sum of those CUSUM statistics is more effective than the maximum when a moderate or large number of data streams are abnormal, and vice versa when only a few data streams are abnormal. In practice, it is usually unknown in advance how many data streams will be abnormal. Therefore, neither the maximum or the sum of those CUSUM statistics as the global monitoring statistic guarantees robust performance. In addition, both approaches are based on the CUSUM statistic. In order to obtain the optimal performance of the CUSUM statistic, users need to know the post-change mean.  However, in practice, it is rarely known in advance what kind of changes will occur. If the post-change mean is misspecified, the detection power of the CUSUM statistic will be compromised. Xie and Siegmund (2013) recognized those limitations and proposed a global monitoring statistic derived from the likelihood function of a normal-mixture model. Besides being computationally intensive, their approach also needs to pre-specify the percentage of abnormal data streams. If the percentage is misspecified, their approach will sacrifice some power. Zou et al. (2015) proposed some alternative way to combine the CUSUM statistics from all data streams to produce a single global monitoring statistic. This approach does not require prior knowledge about how many data streams are abnormal and performs well compared to the aforementioned methods across different abnormal scenarios. However, this approach is still based on the CUSUM statistic. Hence it suffers the same drawback as the CUSUM statistic mentioned above. Recently Liu, Zhang and Mei (2017) proposed several global monitoring statistics via the so-called SUM-shrinkage technique. The SUM-shrinkage technique they use is essentially a thresholding method. Instead of taking the sum of all the local monitoring statistics as proposed in Mei (2010), their new global monitoring statistics take the sum of only those that exceed some pre-specified threshold. As expected, the detection power of their monitoring schemes depends on the pre-specified threshold. In their paper, they studied several choices of threshold. However, each one works well only for certain types of abnormal scenarios. Similar to other thresholding-based methods, it is impossible to find a single threshold in their proposed global monitoring statistics that will work well for different types of abnormal scenarios.

In addition to the limitations described above, most of the existing approaches were developed under the normality assumption. When this normality assumption does not hold, none of the above approaches will perform as expected. So all this indicates the need to develop flexible global monitoring statistics that can work with any type of data and be effective for different types of abnormal scenarios. To meet this need, we propose a novel class of global monitoring statistics by making use of the quantile information in the underlying distribution of the local monitoring statistic. The unique use of the quantile information makes our proposed monitoring statistics efficient for different abnormal scenarios, as shown in our simulation studies. Our proposed global monitoring statistics are easy to calculate and can work under flexible model assumptions since they can be built on any local monitoring statistic that is suitable for monitoring a single data stream. There is vast literature on how to monitor a single data stream. Therefore, the proposed class of global monitoring statistics can easily benefit from this rich literature.

The rest of the paper is organized as follows. In Section 2, we propose a general class of global monitoring statistics and show that the global monitoring statistic studied in Zou et al. (2015) can be considered as a special case of our proposed global monitoring statistic. In Section 3, we give three examples of our proposed global monitoring statistic from the general class and evaluate their performance through simulation studies. Finally, we provide some concluding remarks in Section 4.

\section{Methodology}\label{sec:Method}

\subsection{Notation}
The setup for our high-dimension data monitoring problem is the following. There are $m$ data streams in the system.  We denote the observation from the $i$-th data stream at time $t$ by $X_{i,t}$, $i=1,...,m$, $t=1,2,...$. Since a time series model can be used to decorrelate the temporal correlation within each data stream and a spatial model can be used to decorrelate the spatial correlation between data streams before applying monitoring schemes, without loss of generality, we assume that the $X_{i,t}$ are independent both within and between data streams. When the system is in-control (IC), the underlying distribution of $\{X_{i,1},X_{i,2}, ...\}$  ($i=1,...,m$) is called the IC distribution, denoted by $F_{0,i}$. Following this setup, at a given time $t$, we observe $X_{i,1}, X_{i,2}, ..., X_{i,t}$, $i=1,...,m$. The task of our online monitoring scheme at time $t$ is to determine if the distribution of $X_{i,1}, X_{i,2}, ..., X_{i,t}$ is the same as $F_{0,i}$ for all $i=1,2, ...,m$.

The above task can be carried out by tracking a global monitoring statistic $G_t$, which contains information collected from all data streams up to time $t$. If $G_t$ is within the pre-set control limit, we will declare that all data streams are IC and continue monitoring. If $G_t$ exceeds the control limit, we will raise an alarm suggesting some of the data streams are out-of-control (OC). In the following, we propose a novel class of global monitoring statistics that can work with any type of data and be effective for different types of OC scenarios.

\subsection{Proposed global monitoring statistics}

Because the change-point can happen at different times for different data streams, a popular approach in the literature for developing the global monitoring statistic $G_t$ is to first choose an appropriate local monitoring statistic for tracking each data stream and then combine those local monitoring statistics in a way that produces a single global monitoring statistic. We will follow this approach. More specifically, let $W_{i,t}$ be the local monitoring statistic for the $i$-th data stream at time  $t$ that summarizes the evidence regarding a possible local change based on the observations, $X_{i,1}, ..., X_{i,t}$. Without loss of generality, we assume that a larger $W_{i,t}$ indicates a higher probability of the $i$-th data stream being OC. Although our proposed global monitoring statistic $G_t$ can work with any choice of $W_{i,t}$, in order for $G_t$ to be efficient for detecting changes in any data stream,  the $W_{i,t}$ should be chosen to be efficient for detecting local changes. Since choosing a good local monitoring statistic $W_{i,t}$ is equivalent to choosing an appropriate monitoring statistic for the univariate data stream, there is  rich literature on this topic (see, for example, Qiu (2014)) and we can easily find the appropriate monitoring statistic from the literature as $W_{i,t}$ for any particular application in mind. Therefore, in the following, we assume that the $W_{i,t}$ have been constructed, and our focus is how to combine these local monitoring statistics $W_{i,t}$ into a powerful global monitoring statistic.

Note that at any time $t$, we have calculated $W_{1,t},...,W_{m,t}$.  Without loss of generality, we assume that the $W_{i,t}$ are independent and identically distributed when the system is IC. As mentioned in the Introduction, Liu, Zhang and Mei (2017) recently proposed a SUM-shrinkage approach to construct the global monitoring statistic based on $W_{1,t},...,W_{m,t}$. In their approach, $W_{1,t},...,W_{m,t}$ are compared with some pre-specified threshold, and only those that exceed the threshold are used to construct the global test statistic. However, similar to all the other thresholding methods, it is impossible to choose a threshold in advance that works well for all OC scenarios. Instead of comparing $W_{1,t},...,W_{m,t}$ with some pre-specified threshold, we propose to compare their order statistics with their respective expected values when the system is IC. More specifically, let $W_{(1),t} \leq W_{(2),t}\leq \cdots \leq W_{(m),t}$ be the order statistics of $W_{1,t},...,W_{m,t}$. Then  $W_{(i),t}$ is the observed $(i-3/4)/(m-1/2)$ quantile of the underlying distribution of the $W_{i,t}$. Here  $(i-3/4)/(m-1/2)$ is the common continuity correction of $i/m$.  Let $q_{(i),t}$ denote the expected $(i-3/4)/(m-1/2)$ quantile of the IC distribution of the $W_{i,t}$. According to the large-sample asymptotics, when the system is IC, we have
\[
\sqrt{m}\left\{\begin{pmatrix}
W_{(1),t}\\
\vdots\\
W_{(m),t}
\end{pmatrix}
-\begin{pmatrix}
q_{(1),t}\\
\vdots\\
q_{(m),t}
\end{pmatrix}  \right\} \overset{d}{\longrightarrow}N_m \left\{\begin{pmatrix}
0\\
\vdots\\
0
\end{pmatrix},
\begin{pmatrix}
\frac{p_1(1-p_1)}{f(q_{(1),t})^2} & \cdots & \frac{p_1(1-p_m)}{f(q_{(1),t})f(q_{(m),t})}\\
\vdots & \cdots& \vdots\\
 \frac{p_1(1-p_m)}{f(q_{(1),t})f(q_{(m),t})}    & \cdots & \frac{p_m(1-p_m)}{f(q_{(m),t})^2}
\end{pmatrix} \right\},
\]
where ``$\overset{d}{\longrightarrow}$'' stands for ``converges in distribution'', $N_k(\mbs{\mu},\Sigma)$ stands for the $k$-dimensional multivariate normal distribution with mean vector $\mbs{\mu}$ and covariance matrix $\Sigma$,  $p_i=(i-3/4)/(m-1/2)$, and $f(\cdot)$ is  the probability density function of the IC distribution of the $W_{i,t}$.

Ignoring the correlations between the $W_{(i),t}$, a natural statistic which summarizes the differences between the observed quantiles and expected quantiles and takes into account the variabilities of the observed quantiles is
\[
\sum_{i=1}^m \frac{f(q_{(i),t})^2}{p_i(1-p_i)} \omega(i) \Big(W_{(i),t}-q_{(i),t}\Big)^2,
\]
where $\omega(i)$ is the weight function. To make the above statistic more sensitive to the extreme quantiles, we can choose $\omega(i)$ such that it achieves large values for small and large $i$. One of such choices is $\omega(i)=p_i(1-p_i)/f(q_{(i),t})^2$, since $\omega(i)=p_i(1-p_i)/f(q_{(i),t})^2$ usually takes large values for small and large $i$. Using this weight function, a global monitoring statistic that can summarize the differences between the observed quantiles and expected quantiles of the $W_{i,t}$ with emphasis on the extreme quantiles is simply $\sum_{i=1}^m \Big(W_{(i),t}- q_{(i),t} \Big)^2$. Since a larger $W_{i,t}$ indicates a higher probability of the $i$-th data stream being OC,  only when $W_{(i),t}$ is larger than its expected value $q_{(i),t}$, it may indicate abnormality in the system. Therefore,
we only include the difference when $W_{(i),t}$ is larger than its expected value $q_{(i),t}$ in our global monitoring statistic, and the new class of global monitoring statistics we propose is
\begin{equation}
\label{eqn:Gt}
G_t=\sum_{i=1}^m \Big(W_{(i),t}- q_{(i),t} \Big)^2 I_{\{W_{(i),t}>q_{(i),t}\}},
\end{equation}
where $I_{\{A\}}$ is the indicator function and takes 1 if $A$ is true and 0 otherwise. Then our proposed monitoring scheme is to plot $G_t$ over the time $t$ and it raises an alarm if $G_t>h$, where $h$ is the control limit predetermined by the desired IC average run length (denoted by $\text{ARL}_0$).

As seen above, the proposed new class of global monitoring statistics is very general and can work with any local monitoring statistic $W_{i,t}$ that is suitable for the particular application in mind. Therefore, it can find wide-ranging applications in the real world and offer promising solutions to
various statistical process monitoring problems. As shown in our simulation studies reported in Section 3, this class of global monitoring statistics performs very well across different OC scenarios under different model assumptions.

To calculate the above $G_t$, the expected quantiles $q_{(i),t}$ ($i=1,...,m$) of the IC distribution of the $W_{i,t}$ are needed. When the IC distribution of the $W_{i,t}$ is from some well-known distribution family, the $q_{(i),t}$ can be easily obtained from that distribution family. When the IC distribution of the $W_{i,t}$ is not from any well-known distribution family, which is most often the case, we can easily obtain the approximations of the $q_{(i),t}$ through Monte-Carlo simulation. We will give several examples on how to obtain those approximations in the next section. It should be noted that obtaining approximations of the $q_{(i),t}$ will be carried out offline before the online monitoring starts and the values will be stored beforehand. Therefore the total online computational effort in calculating $G_t$ is not different from calculating $\sum_{i=1}^m \Big(W_{(i),t}- a_{i} \Big)^2 I_{\{W_{(i),t}>a_{i}\}}$ with all the $a_{i}$ given.

\subsection{A special case: the global monitoring statistic proposed by Zou et al. (2015)}
Assume that all the IC distributions $F_{0,i}$ are the normal distribution with mean $0$ and variance 1 (denoted by $N(0,1)$), and the OC distribution of the $i$-th data stream ($i=1,...,m$) is also some normal distribution with mean $\mu_{i}$  and variance $1$ (denoted by $N(\mu_{i},1)$). Under those assumptions, an optimal local monitoring statistic for each data stream is the CUSUM statistic. To detect a positive mean shift, the CUSUM statistic for the $i$-th data stream is defined as
\begin{equation}
\label{eqn:CUSUM}
\begin{cases}
S_{i,0}^+=0,\\
S_{i,t}^+=\max(0,S_{i,t-1}^++\mu_i(X_{i,t}-\frac{1}{2}\mu_{i})), \text{ for } t \geq 1.
\end{cases}
\end{equation}
Let $H_{i,t}(\cdot)$ denote the cumulative distribution function of $S^{+}_{i,t}$ when the $i$-th data stream is IC. Then define $U_{i,t}=H_{i,t}(S^+_{i,t})$, $i=1,...,m$, and their order statistics are $U_{(1),t}\leq \cdots \leq U_{(m),t}$. Utilizing one of the goodness-of-fit test statistics developed by Zhang (2002), Zou et al. (2015) proposed the following global monitoring statistic,
\begin{equation}
\label{eqn:GtZ}
G^{Z}_t=\sum_{i=1}^m\Big\{\log\Big[\frac{U_{(i),t}^{-1}-1}{(m-1/2)/(i-3/4)-1} \Big] \Big\}^2I_{\{U_{(i),t}>(i-3/4)/(m-1/2)\}}.
\end{equation}

In the following, we show that $G_t^Z$ can be considered as a special case of our proposed global monitoring statistic $G_t$ in (\ref{eqn:Gt}). To see this, note that $Q(p)=-\log(p^{-1}-1)$ is the quantile function of the standard logistic distribution. If we assume that the OC means $\mu_i$ , $i=1,...,m$, are all equal, then $-\log(U_{(i),t}^{-1}-1)$ can be considered as the observed $(i-3/4)/(m-1/2)$ quantile of the underlying distribution of $S^+_{i,t}$ on the scale of the standard logistic distribution. Similarly, $-\log(\left[(i-3/4)/(m-1/2)\right]^{-1}-1)$ is the expected $(i-3/4)/(m-1/2)$ quantile of the underlying distribution of $S^+_{i,t}$ on the scale of the standard logistic distribution. As a result, if we choose $W_{i,t}$ to be $-\log(U_{i,t}^{-1}-1)$ with $W_{(i),t}=-\log(U_{(i),t}^{-1}-1)$ and $q_{(i),t}=-\log(\left[(i-3/4)/(m-1/2)\right]^{-1}-1)$, our proposed global monitoring statistic $G_t$ in (\ref{eqn:Gt}) reduces to $G_t^Z$ in (\ref{eqn:GtZ}).

The above shows that $G_t^Z$ is a special case of our proposed global monitoring statistic $G_t$. Therefore, $G_t^Z$ can be also considered as the sum of the squared differences between the observed quantiles and expected quantiles. Theoretically $G_t^Z$ can be modified using local monitoring statistics other than the above CUSUM statistics $S^+_{i,t}$. However, the quantiles used in $G_t^Z$ are on the scale of the standard logistic distribution. To obtain those quantiles, the local monitoring statistics have to be transformed to $U_{i,t}$ based on their underlying IC distribution. For many commonly used local monitoring statistics, there is no analytical form available for this transformation and it has to be approximated through the Markov chain method or Monte Carlo simulation, which can be time-consuming, especially for high-dimensional data monitoring. This greatly restricts the applicability of $G_t^Z$ to other settings. Therefore, in Zou et al. (2015), all the analysis was limited to using $S^+_{i,t}$ in (\ref{eqn:CUSUM}) as the local monitoring statistic, since a close-form formula to approximate $U_{i,t}$ is available in this setting thanks to Grigg and Spiegelhalter (2008). In contrast, the quantiles used in our proposed $G_t$ can be directly determined by any local monitoring statistics $W_{i,t}$, which makes our $G_t$ more versatile and more computationally efficient than $G_t^Z$ for monitoring high-dimensional data.

\section{Examples}
In Section \ref{sec:Method}, we propose a general class of global monitoring statistics, which can be built on any local monitoring statistic. In this section, we provide three examples of our proposed global monitoring statistic from this general class and compare their performance with that of other existing global monitoring statistics.

\subsection{Known pre- and post-change distributions} \label{sec:ex1}

In our first example, we assume that the distributions before and after the change are $N(0,1)$ and $N(\mu,1)$, respectively, for all the data streams, where $\mu$ is the post-change mean and is completely specified. Under this setting, we use the CUSUM statistic $S^+_{i,t}$ defined in (\ref{eqn:CUSUM}) with $\mu_i=\mu$ as the local monitoring statistic. 

Based on this local monitoring statistic, the global monitoring statistic $G_t^Z$ defined in (\ref{eqn:GtZ}) can be used to monitor the $m$ data streams jointly. As mentioned earlier, to implement $G_t^Z$, it is important that $U_{i,t}$ can be calculated quickly. Grigg and Spiegelhalter (2008) developed an empirical approximation to the IC steady-state distribution of the CUSUM statisitc $S^+_{i,t}$. Their result can be used to obtain a close-form formula to calculate $U_{i,t}$. Since this formula only works when $S^+_{i,t}$ reaches its steady-state, to make use of this formula, we modify the definition of the CUSUM statistic a little. Instead of starting the CUSUM statistic at 0, i.e., $S_{i,0}^+=0$ as in (\ref{eqn:CUSUM}), we start the CUSUM statistic at some value randomly drawn from the IC steady-state distribution of $S_{i,t}^+$. More specifically, we first generate $10^5$ independent sequences of $\{X_{k,1},...,X_{k,2000}\}$ ($k=1,...,10^5$), each of which is independently drawn from $N(0,1)$, and calculate $S_{k,2000}^+$ as in (\ref{eqn:CUSUM}). Then $\{S_{k,2000}^+\}_{k=1}^{10^5}$ can serve as a random sample from the IC steady-state distribution of $S_{i,t}^+$. Our modified CUSUM statistic is then  defined as follows. For $i=1,...,m$,
\begin{equation}
\label{eqn:CUSUM1}
\begin{cases}
S_{i,0}^{+*}=V_i\\
S_{i,t}^{+*}=\max(0,S_{i,t-1}^{+*}+\mu(X_{i,t}-\frac{1}{2}\mu)), \text{ for } t \geq 1
\end{cases}
\end{equation}
where $V_i$ is randomly drawn with replacement from  $\{S_{k,2000}^+\}_{k=1}^{10^5}$. The global monitoring statistic $G_t^Z$ in (\ref{eqn:GtZ}) is then calculated using $U_{i,t}=H^*(S_{i,t}^{+*})$, where $H^*(\cdot)$ is the IC distribution  of $S_{i,t}^{+*}$. Since $S_{i,t}^{+*}$ starts from the steady-state, $H^*(\cdot)$ at any time $t$ follows the IC steady-state distribution. As a result, we can utilize the close-form formula provided in Grigg and Spiegelhalter (2008) to calculate the above $U_{i,t}$ quickly.

Similarly, we also use the above modified CUSUM statistic $S_{i,t}^{+*}$ as $W_{i,t}$ in our proposed global monitoring statistic $G_t$ in (\ref{eqn:Gt}). Due to this modification, the underlying distribution of $W_{i,t}$ for any time $t$ is the same as that of $\{S_{k,2000}^+\}_{k=1}^{10^5}$ obtained above. Then its expected quantiles $q_{(i),t}$ also remain the same for any time $t$, and can be well approximated by the corresponding sample quantiles of $\{S_{k,2000}^+\}_{k=1}^{10^5}$, which we denote by $\hat{q}^s_{(i)}$. Therefore, our proposed global monitoring statistic $G_t$ in this particular setting is
\[
G_t=\sum_{i=1}^m \Big(S_{(i),t}^{+*}- \hat{q}^s_{(i)} \Big)^2 I_{\{S_{(i),t}^{+*}>\hat{q}^s_{(i)}\}},
\]
where $S_{(1),t}^{+*} \leq \cdots \leq S_{(m),t}^{+*}$ are the order statistics of $S_{1,t}^{+*}, ..., S_{m,t}^{+*}$.

In Liu, Zhang and Mei (2017),  several global monitoring statistics based on hard-thresholding, soft-thresholding and order-thresholding were proposed. From their simulations, the soft-thresholding method seems to work the best. Therefore, we only include their soft-thresholding based global monitoring statistic in the following simulation study for performance comparison. To be consistent with the above $G_t^Z$ and $G_t$, we also calculate their global monitoring statistic based on the above modified CUSUM statistic $S_{i,t}^{+*}$, which is defined as
\[
G_t^L=\sum_{i=1}^m \max\{S^{+*}_{i,t}-b,0\},
\]
where $b$ is the thresholding constant. Following Liu, Zhang and Mei (2017),  three choices of $b$ are considered: (i) $b_1=1/2$; (ii) $b_2=\log(10)=2.3026$; (iii) $b_3=\log(100)=4.6052$.

\noindent \textbf{$\bullet$ Simulation study}

In the following, we report a simulation study to compare the performance of $G_t$, $G_t^Z$ and $G_t^L$. The general simulation settings are the following. Among the $m$ data streams, $m_0$ data streams are from the IC distribution $N(0,1)$, and the remaining $m_1=m-m_0$ data streams are from the OC distribution $N(0.5,1)$. We consider two choices of $m$: $m=100$ and $1000$, and Table \ref{tab:ARL1} lists the corresponding choices of $m_1$ for these two choices of $m$.

In our simulation study, we construct the monitoring scheme by tracking $G_t$, $G_t^Z$ and $G_t^L$, respectively. If $G_t$, $G_t^Z$ or $G_t^L$ exceed its respective control limit $h$, the corresponding monitoring scheme will stop the monitoring and raise an alarm. The control limit $h$ for $G_t$, $G_t^Z$ and $G_t^L$ can be obtained through Monte-Carlo simulation to satisfy the $\text{ARL}_0$ requirement. The desired $\text{ARL}_0$ for all the monitoring schemes is set at 1000. The control limits $h$ for  $G_t$, $G_t^Z$ and $G_t^L$ for different values of $m$ are listed in Table \ref{tab:CL1}.
\begin{table}[!hbtp]
\begin{center}
\caption{The control limits of the monitoring schemes based on $G_t$, $G_t^Z$ and $G_t^L$ when $\text{ARL}_0=1000$.}\label{tab:CL1}
\begin{tabular}{|c|c|c|ccc||c|c|ccc|}
  \hline
&\multicolumn{5}{|c||}{ $m=100$} & \multicolumn{5}{|c|}{$m=1000$} \\
\cline{2-11}
& & & \multicolumn{3}{c||}{ $G_t^L$ } & & & \multicolumn{3}{|c|}{ $G_t^L$ }\\
\cline{4-6}
\cline{9-11}
&$G_t$ & $G_t^Z$  & $b_1$ & $b_2$ & $b_3$ &  $G_t$ & $G_t^Z$  & $b_1$ & $b_2$ & $b_3$  \\
\hline
$h$ & 20.674 & 28.570 & 69.496 & 19.303 & 5.513 & 25.13 & 32.593 & 526.599 & 108.212 & 19.413\\
  \hline
\end{tabular}
\end{center}
\end{table}

Using those control limits, the monitoring schemes based on $G_t$, $G_t^Z$ and $G_t^L$ are then used to monitor the above $m$ data streams with $m_1$ of them being OC. Since those OC data streams have changed from their IC distributions from the very beginning, the detection power of the  monitoring schemes based on $G_t$, $G_t^Z$ and $G_t^L$ can be compared by the average time for the monitoring scheme to raise an alarm, i.e., the average run length (denoted by $\text{ARL}_1$). Table \ref{tab:ARL1}
reports the $\text{ARL}_1$ of the monitoring schemes based on $G_t$, $G_t^Z$ and $G_t^L$ for different settings from 2500 simulations. The standard deviations of the run lengths from the 2500 simulations are also included in parentheses, and the standard errors of the $\text{ARL}_1$ are simply those standard deviations divided by 50.
\begin{table}[!hbtp]
\begin{center}
\caption{The $\text{ARL}_1$ of the monitoring schemes based on $G_t$, $G_t^Z$ and $G_t^L$ from 2500 simulations. The standard deviations of the run lengths from the 2500 simulations are reported in parentheses.}
\label{tab:ARL1}
\begin{tabular}{|c|c||c|c||ccc|}
  \hline
  & & & & \multicolumn{3}{c|}{$G_t^L$} \\
  \cline{5-7}
$m$ & $m_1$ & $G_t$ & $G_t^Z$  & $b_1$ & $b_2$ & $b_3$\\
  \hline
& 1 & 63.67 (31.97) & 68.04 (33.15) & 110.26 (55.02) & 81.22 (40.72) & \textbf{62.71} (31.84) \\
& 3 & 36.04 (14.38) & 38.17 (15.01) & 48.44 (21.88) & 38.23 (15.92) & \textbf{35.74} (13.62) \\
& 5 & 27.27 (10.35) & 28.75 (10.96) & 32.55 (14.35) & \textbf{27.26} (10.53) & 28.65 (9.89) \\
& 8 & 20.04 (7.19) & 20.77 (7.76) & 21.19 (8.94) & \textbf{19.67} (7.06) & 23.04 (7.53) \\
100 & 10 & 17.32 (6.23) & 17.63 (6.64) & 17.33 (7.21) & \textbf{17.01} (5.93) & 21.08 (6.52) \\
& 20 & 10.65 (3.63) & 10.04 (3.84) & \textbf{9.43} (3.61) & 11.25 (3.52) & 16.12 (4.56) \\
& 50 & 4.89 (1.56) & 3.91 (1.43) & \textbf{4.15} (1.40) & 6.52 (1.97) & 11.23 (3.33) \\
& 80 & 3.25 (1.00) & 2.37 (0.82) & \textbf{2.81} (0.93) & 4.99 (1.54) & 9.48 (2.81) \\
& 100 & 2.68 (0.78) & 1.89 (0.59) & \textbf{2.37} (0.74) & 4.32 (1.34) & 8.56 (2.63) \\
    \hline
    \hline
& 1 & 82.18 (37.20) & 86.87 (38.64) & 214.47 (113.31) & 148.94 (74.62) & \textbf{98.34} (44.36) \\
& 3 & 51.55 (19.03) & 53.97 (19.81) & 106.37 (49.85) & 73.70 (31.93) & \textbf{53.60} (20.03) \\
& 5 & 41.57 (13.72) & 43.36 (14.39) & 73.63 (32.83) & 52.33 (20.96) & \textbf{41.54} (13.64) \\
& 8 & 34.01 (10.58) & 35.47 (11.02) & 52.97 (21.97) & 38.89 (14.17) & \textbf{33.70} (9.99) \\
& 10 & 30.45 (9.27) & 31.58 (9.63) & 43.87 (17.80) & 33.02 (11.62) & \textbf{30.07} (8.80) \\
& 20 & 21.31 (6.24) & 21.60 (6.60) & 24.74 (9.58) & \textbf{20.88} (6.69) & 22.54 (5.85) \\
& 50 & 12.10 (3.57) & 11.50 (3.88) & \textbf{11.01} (4.03) & 11.88 (3.30) & 15.82 (3.70) \\
& 80 & 8.53 (2.52) & 7.65 (2.69) & \textbf{7.18} (2.55) & 8.94 (2.43) & 13.09 (3.09) \\
& 100 & 7.13 (2.10) & 6.16 (2.19) & \textbf{5.83} (2.01) & 7.85 (2.07) & 12.14 (2.71) \\
1000 & 150 & 5.07 (1.48) & 4.08 (1.45) & \textbf{4.09} (1.36) & 6.16 (1.65) & 10.34 (2.46) \\
& 200 & 3.94 (1.14) & 3.05 (1.06) & \textbf{3.20} (1.04) & 5.19 (1.45) & 9.16 (2.23) \\
& 300 & 2.83 (0.77) & 2.07 (0.64) & \textbf{2.35} (0.71) & 4.14 (1.13) & 7.80 (2.00) \\
& 400 & 2.22 (0.60) & 1.61 (0.53) & \textbf{1.92} (0.57) & 3.46 (0.96) & 6.86 (1.82) \\
& 500 & 1.87 (0.46) & 1.30 (0.46) & \textbf{1.65} (0.51) & 3.04 (0.86) & 6.18 (1.68) \\
& 600 & 1.67 (0.48) & 1.08 (0.28) & \textbf{1.45} (0.50) & 2.71 (0.80) & 5.65 (1.57) \\
& 700 & 1.46 (0.50) & 1.02 (0.13) & \textbf{1.28} (0.45) & 2.47 (0.72) & 5.22 (1.50) \\
& 800 & 1.26 (0.44) & 1.00 (0.03) & \textbf{1.16} (0.37) & 2.27 (0.68) & 4.92 (1.40) \\
& 900 & 1.10 (0.30) & 1.00 (0.00) & \textbf{1.07} (0.25) & 2.11 (0.62) & 4.60 (1.32) \\
&1000 & 1.03 (0.17) & 1.00 (0.00) & \textbf{1.03} (0.17) & 1.97 (0.59) & 4.31 (1.32) \\
  \hline
\end{tabular}
\end{center}
\end{table}

For the monitoring schemes based on $G_t^L$, the bold number in each row of Table \ref{tab:ARL1} represents the smallest $\text{ARL}_1$ among the three choices of $b$ for that particular OC scenario. As the table shows, the detection power of $G_t^L$ depends on the choice of $b$. If  $b$ is too small,  $G_t^L$ is not powerful when only a few data streams are OC, since many IC data streams may exceed $b$ and the signal in  $G_t^L$ will be diluted by including those IC data streams. Similarly, if $b$ is too large,   $G_t^L$ is not powerful when many data streams are OC, since many of them may not exceed $b$ and hence do not contribute to $G_t^L$. This is consistent with what is known for any thresholding-based method. However, in practice, it is rarely known in advance how many data streams will be OC, which makes it extremely difficult to come up with an appropriate $b$ for  $G_t^L$ in real applications. In contrast, both $G_t$ and $G_t^Z$ do not depend on any sort of tuning parameter, and they perform well across different OC scenarios with detection delays being always close to those of $G_t^L$ with the best choice of $b$.

When further comparing $G_t$ with $G_t^Z$, we notice that our $G_t$ is better when a small number of data streams are OC, while $G_t^Z$ is better when a large number of data streams are OC. This can be explained by the following. As described in Section 2, both of the global monitoring statistics can be viewed as the sum of the squared differences between the observed quantiles and expected quantiles. $G_t^Z$ is based on the quantiles from the standard logistic distribution, while our $G_t$ is based on the quantiles from the distribution of the CUSUM statistic $S^{+*}_{i,t}$ in (\ref{eqn:CUSUM1}). As shown in Grigg and Spiegelhalter (2008), the tail of the IC distribution of $S^{+*}_{i,t}$ resembles that of an exponential distribution, which implies that the IC distribution of $S^{+*}_{i,t}$ has a heavier tail than the standard logistic distribution. As a result, the extreme quantiles contribute more in $G_t$ than in $G_t^Z$. This explains why $G_t$ performs better than $G_t^Z$ when a small number of data streams are OC, and vice versa when a large number of data streams are OC.

\subsection{Known pre-change distribution but unknown post-change distribution}\label{sec:ex2}

In the previous example, in order to use the CUSUM statistic as the local monitoring statistic, the distribution after the change need to be completely specified. This can be a difficult task for many real-world applications, where prior knowledge of the post-change distribution may not be available. In our second example, we consider a more realistic setting. That is, we assume that the OC distribution of the $i$-th data stream is $N(\mu_i,1)$ with $\mu_i$ unknown, and the IC distributions of all data streams are still $N(0,1)$. To obtain the specific form of our proposed global monitoring statistic $G_t$, the key is to find the appropriate local monitoring statistic $W_{i,t}$ in this setting. There exist a few options for such a statistic in the statistical process monitoring literature. For example, Sparks (2000) proposed an adaptive CUSUM statistic, and Han ad Tsung (2006) developed a reference-free-cumulative-score statistic. In both of the two methods, instead of using the specified $\mu_i$ in the CUSUM statistic defined in (\ref{eqn:CUSUM}), an estimate of $\mu_i$ is plugged in. In Sparks (2000), an exponentially weighted moving average of all the past observations is used to estimate $\mu_i$, while, in Han and Tsung (2006), the absolute value of the current observation, $|X_{i,t}|$, is used as the estimate of $\mu_i$. Following the same idea, Lorden and Pollak (2008) proposed another estimate of $\mu_i$ to replace $\mu_i$ in the CUSUM statistic in (\ref{eqn:CUSUM}), and proved the asymptotic optimality of the resulting monitoring statistic. Since the CUSUM statistic in (\ref{eqn:CUSUM}) is only for detecting positive mean shifts, the monitoring statistic developed in Lorden and Pollak (2008) is also only for positive mean shifts. Recently, Liu, Zhang and Mei (2017) extended Lorden and Pollak's monitoring statistic to detect both positive and negative mean shifts. In the following, we use this two-sided monitoring statistic in Liu, Zhang and Mei (2017) as our local monitoring statistic $W_{i,t}$.

More specifically, define, for $t \geq 1$,
\begin{align*}
C_{i,t}^{(1)}&=\max\Big(0,C_{i,t-1}^{(1)}+\hat{\mu}_{i,t}^{(1)}(X_{i,t}-\frac{1}{2}\hat{\mu}_{i,t}^{(1)})\Big) ,\\
C_{i,t}^{(2)}&=\max\Big(0,C_{i,t-1}^{(2)}+\hat{\mu}_{i,t}^{(2)}(X_{i,t}-\frac{1}{2}\hat{\mu}_{i,t}^{(2)})\Big) ,
\end{align*}
where $\hat{\mu}_{i,t}^{(1)}$ and $\hat{\mu}_{i,t}^{(2)}$ are the estimates of $\mu_i$ for the positive mean shift and negative mean shift, respectively, and they are given by
\[
\hat{\mu}_{i,t}^{(1)}=\max\Big(\rho, \frac{s+S_{i,t}^{(1)}}{t+T_{i,t}^{(1)}} \Big)>0, \quad \hat{\mu}_{i,t}^{(2)}=\min\Big(-\rho, \frac{-s+S_{i,t}^{(2)}}{t+T_{i,t}^{(2)}} \Big)<0.
\]
In the above estimates, $\rho$ is the pre-specified smallest mean shift that is meaningful, and $s$ and $t$ are also pre-specified nonnegative constants and can be considered as a prior so that the above estimates can be treated as the Bayes-type estimates. In our simulation studies, we choose $\rho=0.25$, $s=1$ and $t=4$ as in Liu, Zhang and Mei (2017). For $j=1,2$, the sequences $(S_{i,t}^{(j)}, T_{i,t}^{(j)})$ are calculated recursively
\begin{align*}
S_{i,t}^{(j)}&=\begin{cases}
S_{i,t-1}^{(j)}+X_{i,t-1}, & \text{if } C^{(j)}_{i,t-1}>0, \\
0, &  \text{if } C^{(j)}_{i,t-1}=0,
\end{cases}\\
T_{i,t}^{(j)}&=
\begin{cases}
T_{i,t-1}^{(j)}+1, & \text{if } C^{(j)}_{i,t-1}>0, \\
0, &  \text{if } C^{(j)}_{i,t-1}=0.
\end{cases}
\end{align*}

Finally, our local monitoring statistic $C_{i,t}$ is simply
\[
C_{i,t}=\max(C_{i,t}^{(1)},C_{i,t}^{(2)}).
\]

In Liu, Zhang and Mei (2017), the above monitoring statistic $C_{i,t}$ starts from the following initial values:
\[
S_{i,0}^{(1)}=S_{i,0}^{(2)}=T_{i,0}^{(1)}=T_{i,0}^{(2)}=C_{i,0}^{(1)}=C_{i,0}^{(2)}=X_{i,0}=0.
\]
Using those initial values, the IC distribution of the $C_{i,t}$ will change over the time before it reaches its steady-state. Recall that, to implement our proposed global monitoring statistic $G_t$, the expected quantiles $q_{(i),t}$ of the IC distribution of the $C_{i,t}$ are needed. If the IC distribution of the $C_{i,t}$ changes over the time, then we need to calculate and store $\{q_{(i),t}\}_{i=1}^m$ for each $t$. To simplify our procedure, similarly to how we modified the original CUSUM statistic in the previous section, we propose to set the initial values, $(S_{i,0}^{(1)},S_{i,0}^{(2)},T_{i,0}^{(1)},T_{i,0}^{(2)},C_{i,0}^{(1)},C_{i,0}^{(2)},X_{i,0})$,
at some value randomly drawn from the IC steady-state distribution of $(S_{i,t}^{(1)},S_{i,t}^{(2)},T_{i,t}^{(1)},T_{i,t}^{(2)},C_{i,t}^{(1)},C_{i,t}^{(2)},X_{i,t})
$. To obtain such initial values, we generate $10^5$ independent sequences of $\{X_{k,1},...,X_{k,2000}\}$ ($k=1,...,10^5$), each of which is independently drawn from $N(0,1)$, and calculate $(S_{k,2000}^{(1)},S_{k,2000}^{(2)},T_{k,2000}^{(1)},T_{k,2000}^{(2)},C_{k,2000}^{(1)},C_{k,2000}^{(2)},X_{k,2000})$,
using the initial values $0$. Then $\{(S_{k,2000}^{(1)},S_{k,2000}^{(2)},T_{k,2000}^{(1)},T_{k,2000}^{(2)},C_{k,2000}^{(1)},C_{k,2000}^{(2)},X_{k,2000})\}_{k=1}^{10^5}$
can be used to approximate the IC steady-state distribution of $(S_{i,t}^{(1)},S_{i,t}^{(2)},T_{i,t}^{(1)},T_{i,t}^{(2)},C_{i,t}^{(1)},C_{i,t}^{(2)},X_{i,t})
$. The initial values to calculate our modified $C^*_{i,t}$ are then defined as,
\[
(S_{i,0}^{(1)},S_{i,0}^{(2)},T_{i,0}^{(1)},T_{i,0}^{(2)},C_{i,0}^{(1)},C_{i,0}^{(2)},X_{i,0})=\mbs{V}_i,
\]
where $\mbs{V}_i$ is randomly drawn  from $\{(S_{k,2000}^{(1)},S_{k,2000}^{(2)},T_{k,2000}^{(1)},T_{k,2000}^{(2)},C_{k,2000}^{(1)},C_{k,2000}^{(2)},X_{k,2000})\}_{k=1}^{10^5}$ with replacement. Since $(S_{i,t}^{(1)},S_{i,t}^{(2)},T_{i,t}^{(1)},T_{i,t}^{(2)},C_{i,t}^{(1)},C_{i,t}^{(2)},X_{i,t})
$ starts from the steady-state, $C^*_{i,t}$ at any time $t$ follows the IC steady-state distribution when the system is IC, and its expected quantiles $q_{(i),t}$ also remain the same for any time $t$. Then $\{\text{max}(C^{(1)}_{k,2000},C^{(2)}_{k,2000})\}_{k=1}^{10^5}$ can be used to approximate the IC steady-state distribution of $C^*_{i,t}$, and the expected quantiles $q_{(i),t}$ of $C^*_{i,t}$ can be approximated by the corresponding sample quantiles of $\{\text{max}(C^{(1)}_{k,2000},C^{(2)}_{k,2000})\}_{k=1}^{10^5}$, which we denote by $\hat{q}^{c}_{(i)}$. Therefore, our proposed global monitoring statistic $G_t$ in this particular setting is
\[
G_t=\sum_{i=1}^m \Big(C^*_{(i),t}-\hat{q}^c_{(i)} \Big)^2 I_{\{C^*_{(i),t}>\hat{q}^c_{(i)}\}},
\]
where $C^*_{(1),t} \leq \cdots \leq C^*_{(m),t}$ are the order statistics of $C^*_{1,t},...,C^*_{m,t}$.

\noindent \textbf{$\bullet$ Simulation study}

Using the above modified local monitoring statistics $C^*_{i,t}$, theoretically it is possible to define the global monitoring statistic $G_t^Z$ proposed by Zou et al. (2015) in this setting accordingly. To implement this $G_t^Z$, it is important to have a close-form formula for the cumulative distribution function of $C^*_{i,t}$. However, it is not easy to develop such a formula for the above $C^*_{i,t}$. Due to this computational difficulty of $G_t^Z$, in our simulation study we only compare our global monitoring statistic $G_t$ defined above with the one proposed in Liu, Zhang and Mei (2017). Their global monitoring statistic based on soft-thresholding in this particular setting is defined as
\[
G_t^L=\sum_{i=1}^m \max\{C^*_{i,t}-b,0\}.
\]
Again following Liu, Zhang and Mei (2017), three choices of $b$ are considered: (i) $b_1=1/2$; (ii) $b_2=\log(10)=2.3026$; (iii) $b_3=\log(100)=4.6052$.

The specific simulation settings are similar to those in Section \ref{sec:ex1}.  Among the $m$ data streams, $m_0$ data streams are from the IC distribution $N(0,1)$, and the remaining $m_1=m-m_0$ data streams are from the OC distribution $N(\mu_i,1)$, $i=1,...,m_1$, where $\mu_i$ is randomly drawn from $\{-0.5, 0.5\}$. We consider two choices of $m$: $m=100$ and $1000$, and Table \ref{tab:ARL2} lists the corresponding choices of $m_1$ for these two choices of $m$.

Similar to the first simulation study reported in Section \ref{sec:ex1}, the performance of $G_t$ and $G_t^L$ is compared based on the $\text{ARL}_1$ of their corresponding monitoring schemes. The desired $\text{ARL}_0$ for the $G_t$- and $G_t^L$-based monitoring schemes is set at 1000. The control limits $h$ for those monitoring schemes, which are obtained through Monte-Carlo simulation, are listed in Table~\ref{tab:CL2}.
\begin{table}[!hbtp]
\begin{center}
\caption{The control limits of the monitoring schemes based on $G_t$ and $G_t^L$ when $\text{ARL}_0=1000$.}\label{tab:CL2}
\begin{tabular}{|c|c|ccc||c|ccc|}
  \hline
&\multicolumn{4}{|c||}{ $m=100$} & \multicolumn{4}{|c|}{$m=1000$} \\
\cline{2-9}
& & \multicolumn{3}{c||}{ $G_t^L$ }  & & \multicolumn{3}{|c|}{ $G_t^L$ }\\
\cline{3-5}
\cline{7-9}
& $G_t$  & $b_1$ & $b_2$ & $b_3$ &  $G_t$ & $b_1$ & $b_2$ & $b_3$  \\
\hline
$h$ & 19.717 & 78.452 & 19.977 & 5.644 & 24.096 & 617.353 & 114.857 & 19.787\\
  \hline
\end{tabular}
\end{center}
\end{table}

Based on those control limits, the $\text{ARL}_1$ of the $G_t$- and $G_t^L$-based  monitoring schemes are obtained from 2500 simulations, which are reported in Table \ref{tab:ARL2}. The standard deviations of the run lengths from the 2500 simulations are also included in parentheses. Again, for the $G_t^L$-based monitoring schemes, the bold number in each row represents the smallest $\text{ARL}_1$ among the three choices of $b$ for that particular OC scenario. As we can see from the table,  the  $\text{ARL}_1$ of $G_t^L$ depends on the choice of $b$. $G_t^L$ with a small $b$ does not perform well when a small number of data streams are OC, while $G_t^L$ with a large $b$ does not perform well when a large number of data streams are OC. The explanation is similar to that we give in Section \ref{sec:ex1}. Since it is rarely known in advance how many data streams will be OC in practice, it is extremely difficult to come up with an appropriate $b$ for  $G_t^L$ in real applications. In contrast, despite the fact that there is no tuning parameter involved, our $G_t$ performs well across different OC scenarios, and its detection delays are always close to those of $G_t^L$ with the best choice of $b$. This makes our $G_t$ particularly appealing in many real-world applications.

\begin{table}[!hbtp]
\begin{center}
\caption{The $\text{ARL}_1$ comparison of the monitoring schemes based on $G_t$ and $G_t^L$. The standard deviations of the run lengths from the 2500 simulations are reported in parentheses.}
\label{tab:ARL2}
\begin{tabular}{|c|c||c||ccc|}
  \hline
  & & & \multicolumn{3}{c|}{$G_t^L$} \\
  \cline{4-6}
$m$ & $m_1$ & $G_t$  & $b_1$ & $b_2$ & $b_3$\\
  \hline
&  1 & 71.05 (37.20) & 120.08 (62.31) & 88.54 (45.55) & \textbf{70.94} (36.97) \\
&  3 & 41.05 (17.58) & 55.64 (25.85) & 44.01 (19.25) & \textbf{40.89} (17.21) \\
&  5 & 31.37 (12.67) & 37.93 (16.37) & \textbf{31.89} (13.03) & 32.13 (12.57) \\
&  8 & 24.51 (9.27) & 26.89 (10.98) & \textbf{24.25} (9.15) & 26.24 (9.69) \\
100&  10 & 21.54 (8.08) & 22.86 (9.24) & \textbf{21.10} (7.84) & 23.86 (8.73) \\
& 20 & 14.43 (4.77) & \textbf{13.49} (4.85) & 14.25 (4.62) & 18.02 (5.83) \\
&  50 & 8.17 (2.26) & \textbf{6.98} (2.13) & 8.59 (2.44) & 12.41 (3.60) \\
&  80 & 6.17 (1.58) & \textbf{5.20} (1.40) & 6.85 (1.81) & 10.39 (2.87) \\
&  100 & 5.40 (1.26) & \textbf{4.51} (1.14) & 6.16 (1.53) & 9.52 (2.53) \\
   \hline
   \hline
&  1 & 91.72 (43.42) & 229.39 (126.38) & 161.70 (83.56) & \textbf{106.88} (51.26) \\
&  3 & 56.75 (22.74) & 114.71 (56.63) & 80.57 (35.88) & \textbf{59.57} (23.93) \\
& 5 & 46.40 (17.00) & 82.56 (37.59) & 59.86 (24.22) & \textbf{47.45} (16.85) \\
&  8 & 37.48 (13.18) & 59.71 (25.92) & 44.29 (16.67) & \textbf{37.83} (12.81) \\
&  10 & 34.05 (11.68) & 50.09 (20.84) & 38.23 (13.94) & \textbf{33.98} (11.23) \\
&  20 & 24.57 (7.74) & 30.05 (11.50) & 25.26 (8.34) & \textbf{25.19} (7.62) \\
&  50 & 15.36 (4.15) & 15.04 (4.95) & \textbf{14.81} (4.08) & 17.24 (4.50) \\
&  80 & 11.77 (3.06) & \textbf{10.55} (3.30) & 11.36 (2.96) & 14.35 (3.55) \\
&  100 & 10.42 (2.67) & \textbf{9.15} (2.78) & 10.15 (2.65) & 13.17 (3.30) \\
1000&  150 & 8.15 (1.96) & \textbf{6.86} (1.94) & 8.19 (1.98) & 11.26 (2.61) \\
&  200 & 6.94 (1.60) & \textbf{5.74} (1.53) & 7.12 (1.68) & 10.14 (2.31) \\
&  300 & 5.49 (1.18) & \textbf{4.48} (1.08) & 5.90 (1.33) & 8.74 (2.00) \\
&  400 & 4.70 (0.97) & \textbf{3.83} (0.89) & 5.20 (1.12) & 7.91 (1.75) \\
&  500 & 4.19 (0.84) & \textbf{3.43} (0.75) & 4.72 (1.00) & 7.28 (1.59) \\
&  600 & 3.78 (0.76) & \textbf{3.12} (0.69) & 4.33 (0.91) & 6.77 (1.48) \\
&  700 & 3.50 (0.68) & \textbf{2.88} (0.60) & 4.09 (0.84) & 6.45 (1.38) \\
&  800 & 3.28 (0.66) & \textbf{2.71} (0.59) & 3.85 (0.84) & 6.13 (1.34) \\
&  900 & 3.10 (0.58) & \textbf{2.61} (0.55) & 3.70 (0.75) & 5.90 (1.25) \\
&  1000 & 2.95 (0.52) & \textbf{2.47} (0.54) & 3.54 (0.74) & 5.67 (1.25) \\
   \hline
\end{tabular}
\end{center}
\end{table}

\subsection{Unknown pre- and post-change distributions}\label{sec:ex3}

In the previous two examples, the pre-change distributions for all data streams are assumed to be completely known and are specified by some particular parametric distribution. In practice, especially for high-dimension data monitoring, it is often not easy to identify the appropriate parametric distributions for all data streams. Therefore, in this example, we do not assume any particular distribution form for either the pre- or post-change distribution. Again to obtain the specific form of our proposed global monitoring statistic $G_t$ in this setting, we need to find the appropriate local monitoring statistic $W_{i,t}$. To deal with the unknown pre-change distribution, a nonparametric monitoring statistic should be used.
To deal with the unknown post-change distribution, we need a nonparametric monitoring statistic that can detect any arbitrary distributional changes. In the literature, to deal with the unknown pre-change distribution, many nonparametric monitoring statistics assume that a large amount of IC reference data generated by the pre-change distribution is available so that certain characteristics of the pre-change distribution can be well estimated. However, in order for the effect of using estimates instead of the true values on the $\text{ARL}_0$ to be negligible, it usually requires a substantial amount of IC reference data. In many real-world applications, it can be very challenging to have such data. Therefore, to find a good candidate for our $W_{i,t}$, we only focus on the nonparametric monitoring statistics that have the self-starting feature.

There are a few nonparametric self-starting monitoring statistics that can detect any arbitrary distributional changes in the literature. For example, Zou and Tsung (2010) proposed an EWMA statistic based on a powerful goodness-of-fit test. However, according to the simulation studies conducted in Ross and Adams (2012), this EWMA statistic is only sensitive in detecting scale increases and is not as powerful as its competitors in detecting other types of distributional changes including location shifts. Ross and Adams (2012) further proposed two monitoring statistics based on the change-point detection (CPD) framework. Their proposed CPD statistics are shown to have better overall performance than Zou and Tsung's EWMA statistics for detecting different distributional changes. However, like most CPD statistics, the computation of their proposed statistics is very intensive, which
makes them very challenging to implement for monitoring high-dimensional data. Recently Li (2017b) proposed a nonparametric self-starting CUSUM statistic that can detect any arbitrary distributional changes. Based on the simulation studies in Li (2017b), the proposed monitoring statistic is not only computationally more efficient than Ross and Adams's CPD statistics, but also has better overall detection power than those CPD statistics. Therefore, in the following, we use the CUSUM statistic proposed in Li (2017b) as our local monitoring statistic $W_{i,t}$.

Assume that, for each data stream, there are $n$ IC reference data, denoted by $X_{i,-n+1}$, ..., $X_{i,0}$, $i=1,...,m$. At time $t\geq 1$, for the $i$-th data stream, we  partition the real line into the following $d$ left-to-right regions,
\[
\hat{A}^{(1)}_{i,t,1}=(-\infty, \hat{q}^{(1)}_{i,t,1}],\, \hat{A}^{(1)}_{i,t,2}=(\hat{q}^{(1)}_{i,t,1},\hat{q}^{(1)}_{i,t,2}],\,...,\, \hat{A}^{(1)}_{i,t,d}=(\hat{q}^{(1)}_{i,t,d-1},\infty),
\]
and the following $d$ center-outward regions,
\begin{align*}
\hat{A}^{(2)}_{i,t,1}&=(\hat{q}^{(2)}_{i,t,d-1}, \hat{q}^{(2)}_{i,t,d+1}],\\
\hat{A}^{(2)}_{i,t,2}&=(\hat{q}^{(2)}_{i,t,d-2},\hat{q}^{(2)}_{i,t,d-1}] \bigcup (\hat{q}^{(2)}_{i,t,d+1},\hat{q}^{(2)}_{i,t,d+2}],\\
& \cdots \, \cdots\\
\hat{A}^{(2)}_{i,t,d}&=(-\infty, \hat{q}^{(2)}_{i,t,1}] \bigcup (\hat{q}^{(2)}_{i,t,2d-1},\infty),
\end{align*}
where $\hat{q}^{(1)}_{i,t,j}$ ($j=1,...,d-1$) and $\hat{q}^{(2)}_{i,t,k}$ ($k=1,...,2d-1$) are the $(j/d)$-th and $(k/(2d))$-th sample quantiles, respectively, from $X_{i,-n+1},...,X_{i,0}, X_{i,1},...,X_{i,t-1}$. For $j=1,...,d$, define
\[
\hat{Y}^{(1)}_{i,t,j}= I(X_{i,t} \in \hat{A}^{(1)}_{i,t,j}), \quad   \quad \hat{Y}^{(2)}_{i,t,j}= I(X_{i,t} \in \hat{A}^{(2)}_{i,t,j}),
\]
and
\[
\hat{Z}^{(1)}_{i,t,j}=\sum_{l=1}^j \hat{Y}^{(1)}_{i,t,l}, \quad  \quad \hat{Z}^{(2)}_{i,t,j}=\sum_{l=1}^j \hat{Y}^{(2)}_{i,t,l}.
\]
For $k_1, k_2=1,2$, we calculate
\begin{align*}
\hat{S}^{(k_1,k_2)}_{i,t}=\max\Big(0,\hat{S}^{(k_1,k_2)}_{t-1}+ &\frac{1}{d}\sum_{j=1}^{d-1} \frac{d^2}{j(d-j)}\Big\{ \hat{Z}^{(k_1)}_{i,t,j}\log\left(\frac{\sum_{l=1}^j \hat{p}^{(k_1,k_2)}_{i,t,l}}{j/d}\right)\\
&+(1-\hat{Z}^{(k_1)}_{i,t,j})\log\left(\frac{1-\sum_{l=1}^j \hat{p}^{(k_1,k_2)}_{i,t,l}}{1-j/d}\right) \Big\}\Big),
\end{align*}
where $\hat{p}^{(k_1,k_2)}_{i,t,l}$ is defined by
\[
\hat{p}^{(k_1,k_2)}_{i,t,l}=\frac{\alpha^{(k_2)}_l+N^{(k_1,k_2)}_{i,t,l}}{\sum_{j=1}^d \alpha^{(k_2)}_j+ N^{(k_1,k_2)}_{i,t}},
\]
and both $N^{(k_1,k_2)}_{i,t}$ and $N^{(k_1,k_2)}_{i,t,l}$ are calculated recursively by
\begin{align*}
N^{(k_1,k_2)}_{i,t}&=\begin{cases}
N^{(k_1,k_2)}_{i,t-1}+1, & \text{if } \hat{S}^{(k_1,k_2)}_{i,t-1}>0, \\
0, &  \text{if } \hat{S}^{(k_1,k_2)}_{i,t-1}=0,
\end{cases}\\
N^{(k_1,k_2)}_{i,t,l}&=
\begin{cases}
N^{(k_1,k_2)}_{i,t-1,l}+ \hat{Y}^{(k_1)}_{i,t-1,l}, & \text{if } \hat{S}^{(k_1,k_2)}_{i,t-1}>0, \\
0, &  \text{if } \hat{S}^{(k_1,k_2)}_{i,t-1}=0.
\end{cases}
\end{align*}
The constants $\{\alpha^{(k_2)}_1,...,\alpha^{(k_2)}_d\}$ ($k_2=1,2$) serve as the parameters of a prior distribution and are chosen as suggested in Li (2017b).
In particular, when using $\alpha^{(1)}_{j}$ in $\hat{S}^{(1,1)}_{i,t}$, the prior indicates a positive location shift, therefore $\hat{S}^{(1,1)}_{i,t}$ is more powerful for detecting positive location shifts. When using $\alpha^{(2)}_{j}$ in $\hat{S}^{(1,2)}_{i,t}$, the prior indicates a negative location shift, so $\hat{S}^{(1,2)}_{i,t}$ is more powerful for detecting negative location shifts. Similarly, when using $\alpha^{(1)}_{j}$ in $\hat{S}^{(2,1)}_{i,t}$, the prior indicates a scale increase, so $\hat{S}^{(2,1)}_{i,t}$ is more powerful for detecting scale increases. When using $\alpha^{(2)}_{j}$ in $\hat{S}^{(2,2)}_{i,t}$, the prior indicates a scale decrease, so $\hat{S}^{(2,2)}_{i,t}$ is more powerful for detecting scale decreases. If we do not have any prior information about what type of changes the process might encounter, our local monitoring statistic is simply
\begin{equation}
\label{eqn:adaptiveCUSUM_loc_scal}
\hat{S}_{i,t}=\max(\hat{S}^{(1,1)}_{i,t},\hat{S}^{(1,2)}_{i,t},\hat{S}^{(2,1)}_{i,t},\hat{S}^{(2,2)}_{i,t}),
\end{equation}
which is efficient to detect any type of distributional changes. Li (2017b) shows that the above monitoring statistic is distribution-free if $n \geq 2d-1$. Following the suggestion in Li (2017b), we choose $d=20$ and $n=40$.

In Li (2017b), the initial values $(\hat{S}^{(k_1,k_2)}_{i,0}, N^{(k_1,k_2)}_{i,0}, N^{(k_1,k_2)}_{i,0,l},\hat{Y}^{(k_1)}_{i,0,l})$, $k_1,k_2=1,2$, $l=1,...,d$ and $i=1,...,m$, for the above monitoring statistic $\hat{S}_{i,t}$ are all set at 0. To simplify the calculation of our proposed global monitoring statistic $G_t$, similarly to how we modified the local monitoring statistics in the previous two examples,  we propose to set the initial values $(\hat{S}^{(k_1,k_2)}_{i,0}, N^{(k_1,k_2)}_{i,0}, N^{(k_1,k_2)}_{i,0,l},\hat{Y}^{(k_1)}_{i,0,l})$ at some values randomly drawn from their IC steady-state distributions. More specifically, using the distribution-free property of $\hat{S}_{i,t}$, we generate $10^5$ independent sequences of $\{X_{k,-39},...,X_{k,0},X_{k,1},...,X_{k,2000}\}$ ($k=1,...,10^5$), each of which is independently drawn from $N(0,1)$, and calculate
\[
\{(\hat{S}^{(k_1,k_2)}_{k,2000}, N^{(k_1,k_2)}_{k,2000}, N^{(k_1,k_2)}_{k,2000,l},\hat{Y}^{(k_1)}_{k,2000,l})\}_{k=1}^{10^5}
\]
using the initial values $0$. Then $\{(\hat{S}^{(k_1,k_2)}_{k,2000}, N^{(k_1,k_2)}_{k,2000}, N^{(k_1,k_2)}_{k,2000,l},\hat{Y}^{(k_1)}_{k,2000,l})\}_{k=1}^{10^5}$ can be used to approximate the IC steady-state distribution of $(\hat{S}^{(k_1,k_2)}_{i,t}, N^{(k_1,k_2)}_{i,t}, N^{(k_1,k_2)}_{i,t,l},\hat{Y}^{(k_1)}_{i,t,l})$. The initial values to calculate our modified $\hat{S}^*_{i,t}$ are then defined as,
\[
(\hat{S}^{(k_1,k_2)}_{i,0}, N^{(k_1,k_2)}_{i,0}, N^{(k_1,k_2)}_{i,0,l},\hat{Y}^{(k_1)}_{i,0,l})=\mbs{V}_i,
\]
where $\mbs{V}_i$ is randomly drawn  from $\{(\hat{S}^{(k_1,k_2)}_{k,2000}, N^{(k_1,k_2)}_{k,2000}, N^{(k_1,k_2)}_{k,2000,l},\hat{Y}^{(k_1)}_{k,2000,l})\}_{k=1}^{10^5}$ with replacement. The expected quantiles $q_{(i),t}$ of $\hat{S}^*_{i,t}$ can then be well approximated by the corresponding sample quantiles of $\{\text{max}(\hat{S}^{(1,1)}_{k,2000},\hat{S}^{(1,2)}_{k,2000},\hat{S}^{(2,1)}_{k,2000},\hat{S}^{(2,2)}_{k,2000})\}_{k=1}^{10^5}$, which we denote by $\hat{q}^{\hat{s}}_{(i)}$. Therefore, our proposed global monitoring statistic $G_t$ is
\[
G_t=\sum_{i=1}^m \Big(\hat{S}^*_{(i),t}-\hat{q}^{\hat{s}}_{(i)} \Big)^2 I_{\{\hat{S}^*_{(i),t}>\hat{q}^{\hat{s}}_{(i)}\}},
\]
where $\hat{S}^*_{(1),t} \leq \cdots \leq \hat{S}^*_{(m),t}$ are the order statistics of $\hat{S}^*_{1,t},...,\hat{S}^*_{m,t}$.

\noindent \textbf{$\bullet$ Simulation study}

Using the above modified local monitoring statistics $\hat{S}^*_{i,t}$, again it is difficult to implement the global monitoring statistic $G_t^Z$ proposed by Zou et al. (2015), since no close-form formula for the cumulative distribution function of $\hat{S}^*_{i,t}$ is available. We can use the thresholding method proposed in Liu, Zhang and Mei (2017) to come up with some alternative global monitoring statistics. However, it is not clear how to choose a sensible threshold. Therefore, in our simulation study we only compare our global monitoring statistic $G_t$ defined above with two other natural competitors $G_t^{max}=\max_{i=1,...,m}\hat{S}^*_{i,t}$ and $G_t^{sum}=\sum_{i=1}^m \hat{S}^*_{i,t}$.

Again in our simulation study, we consider monitoring $m$ data streams.  Since $\hat{S}^*_{i,t}$ is distribution-free, among the $m$ data streams, we randomly select half of the data streams to have $N(0,1)$ as their IC distributions, one-fifth of the data streams to have the $t$ distribution with 2.5 degrees of freedom as their IC distributions, and the remaining data streams to have the lognormal distribution with parameters $\mu=1$ and $\sigma=0.5$ as their IC distributions. For the data generated from the $t$ or lognormal distribution, we also standardize the data so that their IC distributions have mean 0 and standard deviation 1. Among the $m$ data streams, the first $m_0$ data streams follow their IC distributions all the time, and the remaining $m_1=m-m_0$ data streams will experience certain distributional changes from their IC distributions at the change-point $t=100$. Since  $\hat{S}_{i,t}^*$ is capable of detecting any type of distributional changes, starting from the change-point $t=100$, for the $m_1$ data streams that will experience distributional changes, we add 0.5 to the observations from the first $\lceil m_1/2 \rceil$ data streams to introduce the location change, and multiply 1.5 to the observations from the remaining $m_1-\lceil m_1/2 \rceil$ data streams to introduce the scale change. Here $\lceil b \rceil$ is the smallest integer not less than $b$. Similar to the previous two examples,
we consider two choices of $m$: $m=100$ and $1000$, and Table \ref{tab:ARL3} lists the corresponding choices of $m_1$ for these two choices of $m$. The desired $\text{ARL}_0$ for the $G_t$-, $G_t^{max}$- and $G_t^{sum}$-based monitoring schemes is set at 1000. The control limits $h$ for those monitoring schemes, which are obtained through Monte-Carlo simulation, are listed in Table~\ref{tab:CL3}.
\begin{table}[!hbtp]
\begin{center}
\caption{The control limits of the monitoring schemes based on $G_t$, $G_t^{max}$ and $G_t^{sum}$ when $\text{ARL}_0=1000$.}\label{tab:CL3}
\begin{tabular}{|c|ccc||ccc|}
  \hline
&\multicolumn{3}{|c||}{ $m=100$} & \multicolumn{3}{|c|}{$m=1000$} \\
\cline{2-7}
& $G_t$  & $G_t^{max}$ & $G_t^{sum}$ &  $G_t$ & $G_t^{max}$ & $G_t^{sum}$ \\
\hline
$h$ & 144.016 & 26.908 & 524.492 & 171.697 & 33.492 & 4720.355\\
  \hline
\end{tabular}
\end{center}
\end{table}

Based on those control limits, the $\text{ARL}_1$ (after the change-point) of the $G_t$-, $G_t^{max}$- and $G_t^{sum}$-based  monitoring schemes from 2500 simulations are reported in Table \ref{tab:ARL3}. The standard deviations of the run lengths from the 2500 simulations are also included in parentheses. Again the bold number in each row represents the smaller $\text{ARL}_1$ between $G_t^{max}$ and $G_t^{sum}$ for that particular OC scenario. As we can see from the table, $G_t^{max}$ works best when only a few data streams are OC, but does not perform well when a large number of data streams are OC. On the other hand, $G_t^{sum}$ has the best performance when a large number of data streams are OC, but has the worst performance when only a few data streams are OC. In contrast, our $G_t$ performs well across different OC scenarios, and if its detection delay is not the best among all the three monitoring statistics, it is always very close to the best. This provides another example of the robust performance of our proposed global monitoring statistic $G_t$ for detecting different OC scenarios.

\begin{table}[!hbtp]
\begin{center}
\caption{The $\text{ARL}_1$ comparison of the monitoring schemes based on $G_t$, $G_t^{max}$ and $G_t^{sum}$. The standard deviations of the run lengths from the 2500 simulations are reported in parentheses.}
\label{tab:ARL3}
\begin{tabular}{|c|c||c|cc|}
  \hline
 $m$ & $m_1$ & $G_t$  & $G_t^{max}$ & $G_t^{sum}$\\
  \hline
&  1 &  85.89 (52.00) & \textbf{80.51} (51.40) & 181.85 (136.38) \\
&  3 & 53.43 (25.14) & \textbf{57.68} (27.62) & 86.00 (49.69) \\
&  5 & 43.92 (16.91) & \textbf{51.91} (19.98) & 59.84 (29.17) \\
&  8 & 27.70 (9.55) & 35.90 (13.36) & \textbf{32.82} (14.25) \\
100&  10 & 25.17 (8.44) & 34.07 (12.50) & \textbf{28.33} (11.58) \\
& 20 & 17.24 (4.99) & 28.52 (9.47) & \textbf{16.35} (5.64) \\
&  50 & 10.42 (2.65) & 23.45 (6.99) & \textbf{8.74} (2.55) \\
&  80 &   7.73 (1.87) & 20.61 (6.14) & \textbf{6.30} (1.77) \\
&  100 &  6.63 (1.54) & 19.31 (5.63) & \textbf{5.35} (1.45) \\
   \hline
   \hline
&  1 & 123.93 (118.56) & \textbf{111.90} (98.57) & 452.76 (430.52) \\
&  3 & 75.18 (32.13) & \textbf{76.72} (32.95) & 215.51 (161.58) \\
& 5 & 64.98 (25.32) & \textbf{70.05} (28.96) & 157.96 (104.68) \\
&  8 & 55.19 (20.56) & \textbf{64.05} (25.55) & 112.17 (66.04) \\
&  10 & 50.35 (17.97) & \textbf{61.10} (23.28) & 95.71 (53.16) \\
&  20 & 31.34 (9.16) & \textbf{42.46} (14.31) & 44.15 (19.04) \\
& 50 & 20.66 (5.10) & 34.72 (10.35) & \textbf{21.51} (7.33) \\
&  80 & 15.76 (3.69) & 30.78 (8.23) & \textbf{14.60} (4.50) \\
&  100 & 13.33 (3.01) & 28.14 (7.46) & \textbf{11.82} (3.47) \\
1000&  150 & 10.47 (2.30) & 25.59 (6.49) & \textbf{8.84} (2.48) \\
&  200 & 8.78 (1.90) & 24.14 (6.04) & \textbf{7.14} (1.95) \\
&  300 & 6.83 (1.41) & 22.23 (5.40) & \textbf{5.37} (1.41) \\
&  400 & 5.67 (1.20) & 21.00 (4.90) & \textbf{4.39} (1.17) \\
&  500 & 4.90 (0.99) & 19.87 (4.92) & \textbf{3.79} (0.98) \\
&  600 & 4.45 (0.92) & 19.30 (4.78) & \textbf{3.40} (0.91) \\
&  700 &  4.03 (0.82) & 18.81 (4.60) & \textbf{3.06} (0.80) \\
&  800 & 3.66 (0.75) & 18.23 (4.44) & \textbf{2.76} (0.75) \\
&  900 &  3.37 (0.72) & 17.70 (4.33) & \textbf{2.53} (0.71) \\
&  1000 & 3.12 (0.66) & 17.32 (4.26) & \textbf{2.35} (0.67) \\
   \hline
\end{tabular}
\end{center}
\end{table}

\section{Concluding Remarks}
In this paper, we introduce a general class of global monitoring statistics for high-dimensional data streams. Our proposed global monitoring statistics are easy to calculate and can work with any local monitoring statistic that is suitable for monitoring a single data stream. This flexibility makes our proposed global monitoring statistics suitable for many different real-world applications. To demonstrate the application of this new class of global monitoring statistics, we present three examples and our simulation studies in these three examples further show that our proposed global monitoring statistic performs well under a variety of OC scenarios and has the best overall detection power comparing with other existing global monitoring statistics.


\nocite*{}

\begin{thebibliography}{}


\bibitem[Grigg and Spiegelhalter, 2008]{Grigg2008}
Grigg, O. A., and Spiegelhalter, D. J. (2008).
\newblock An empirical approximation to the null unbounded steady-state distribution of the cumulative sum statistic.
\newblock {\em Technometrics}, \textbf{50}, 501--511.

\bibitem[Han and Tsung, 2006]{Han2006}
Han, D. and Tsung, F. (2006).
\newblock A reference-free cuscore chart for dynamic mean change detection and a unified framework for charting performance comparison.
\newblock {\em Journal of the American Statistical Association}, \textbf{101}, 368-386.


\bibitem[Li, 2017a]{Li2017a}
Li, J. (2017a).
\newblock A two-stage online monitoring procedure for high-dimensional data streams. Accepted by \newblock {\em Journal of Quality Technology}. (arXiv:1712.05074)

\bibitem[Li, 2017b]{Li2017b}
Li, J. (2017b).
\newblock Nonparametric adaptive CUSUM chart for detecting arbitrary distributional changes. \newblock {\em Submitted}. (arXiv:1712.05072)

\bibitem[Liu, 2016]{Liu2016}
Liu, K., Zhang, R. and Mei, Y. (2017).
\newblock Scalable SUM-shrinkage schemes for distributed monitoring
large-scale data streams.
Accepted by \newblock {\em Statistica Sinica}. (arXiv:1603.08652)

\bibitem[Lorden and Pollak, 2008]{Lorden2008}
Lorden, G. and Pollak, M. (2008).
\newblock  Sequential change-point detection procedures that are nearly optimal and computationally simple.
\newblock {\em Sequential Analysis}, \textbf{27}, 476-512.

\bibitem[Mei, 2010]{Mei2010}
Mei, Y. (2010).
\newblock Efficient scalable schemes for monitoring a large number of data streams.
\newblock {\em Biometrika}, \textbf{97}, 419--433.


\bibitem[Qiu, 2014]{Qiu2014}
Qiu, P. (2014). {\it Introduction to statistical process control}, Boca Raton,
FL: Chapman \& Hall/CRC.

\bibitem[Sparks, 2000]{Sparks2000}
Sparks, R. S. (2000).
\newblock CUSUM charts for signalling varying location shifts.
\newblock {\em Journal of Quality Technology}, \textbf{32}, 157--171.


\bibitem[Tartakovsky et al, 2006]{Tartakovsky2006}
Tartakovsky, A. G., Rozovskia, B. L., Blazeka, R. B., and  Kim, H. (2006).
\newblock Detection of intrusions in information
systems by sequential change-point methods (with Discussion).
\newblock {\em Statistical Methodology}, \textbf{3}, 252--340.


\bibitem[Xie and Siegmund, 2013]{Xie2013}
Xie, Y. and Siegmund, D. (2013).
\newblock Sequential multi-sensor change-point detection.
\newblock {\em The Annals of Statistics}, \textbf{41}, 670--692.

\bibitem[Zou et al, 2015]{Zou15}
Zou, C., Jiang, W.,  Wang, Z., and Zi, X.  (2015).
\newblock An efficient on-line monitoring method for high-dimensional data streams.
\newblock {\em Technometrics}, \textbf{57}, 374--387.

\end{thebibliography}

\end{document}